\documentstyle[twoside,fleqn,epsf,espcrc2]{article}
\newcommand{\tkl}{{{\small\sf T}{$\chi$}{\small\sf L}} }

\newcommand{\sesam}{{\small\sf SESAM }}
\newcommand{\sesams}{{\small\sf SESAM}'s }
\newcommand{\sesamc}{{\small\sf SESAM} collaboration}

\newcommand{\beq}{\begin{equation}}
\newcommand{\eeq}{\end{equation}}
\newcommand{\prd}{Phys.~Rev.~D}

\newcommand{\AmS}{{\protect\the\textfont2
  A\kern-.1667em\lower.5ex\hbox{M}\kern-.125emS}}

\title{Bottomonium from NRQCD with Dynamical Wilson Fermions
  \thanks{Talk presented by A. Spitz.}}

\author{\sesam - Collaboration: \\ 
  A.~Spitz$^{\rm a}$,
  N.~Eicker$^{\rm a}$, 
  J.~Fingberg$^{\rm b}$,
  S.~G\"usken$^{\rm b}$, 
  H.~Hoeber$^{\rm b}$, 
  Th.~Lippert$^{\rm a}$, 
  K.~Schilling$^{\rm a,b}$,
  J.~Viehoff$^{\rm b}$. \\[8pt]
  {\small  {\rm $^a$}HLRZ c/o Forschungszentrum J\"ulich, D-52425 J\"ulich,
  and DESY, D-22603 Hamburg, Germany,\\
  {\rm $^b$}Physics Department, University of Wuppertal, D-42097
  Wuppertal, Germany.}}       
\begin{document}
%
%
\begin{abstract}
We present results for the $b \bar b$ spectrum obtained using an 
${\cal O}(M_bv^6)$-correct non-relativis\-tic lattice QCD action.
Propagators are evaluated on \sesams three sets of dynamical gauge
configurations generated with two flavours of Wilson fermions at
$\beta = 5.6$. Compared to a quenched simulation at equivalent lattice
spacing we find better agreement of our dynamical data with 
experimental results in the spin-independent sector but observe no
unquenching effects in hyperfine-splittings. To pin down the
systematic errors we have also compared quenched results in
different ``tadpole'' schemes and used a lower order action. 
\end{abstract}
\maketitle
%
%
\section{INTRODUCTION}
Lattice NRQCD has been widely used in the past few years to calculate a
variety of phenomenologically interesting quantities. The first step in the
NRQCD program, naturally, is the calculation of the spectrum of the $b
\bar b$ system.
In the simulation presented here we use \sesams large sample of
dynamical Wilson-fermion gauge configurations to study both the
spin-independent as well as the spin-dependent spectrum of the
$\Upsilon$. Our strategy, in searching for sea-quark effects, will be
to compare our final dynamical results to that of a quenched
simulation at equivalent lattice spacing.
With our three sea-quark masses we can also study the
dependence of mass-splittings on the light sea-quark mass.
Recent simulations have exposed the sensitivity of spin-independent
splittings to the details of the action. Therefore,
in the spirit of \cite{trottier,manke}, we have implemented
the NRQCD action including spin-dependent corrections of ${\cal O}(M_bv^6)$
and we tadpole improve using the mean link calculated in Landau
gauge. With these ingredients we hope to clarify the effect of
unquenching in the spin-dependent splittings. In addition, using
the quenched configurations we investigate: (i) the
effect of changing tadpole prescriptions; (ii) the effect of using an
${\cal  O}(M_bv^4)$ correct action compared to the ${\cal O}(M_bv^6)$
corrected one. 
%
%
\section{SIMULATION DETAILS}
\subsection{Action}
The non-relativistic (Euclidean) lattice Hamiltonian to ${\cal
  O}(M_bv^6)$ consists of \cite{lepage}: the kinetic energy operator, 
\begin{equation}
  H_0 = -\frac{\Delta^{(2)}}{2M_b} \; ,
\end{equation}
which is of order $M_bv^2$; relativistic corrections of order $M_bv^4$,
\begin{eqnarray}
  \delta H^{(4)} & = & -c_1\frac{\left(\Delta^{(2)}\right)^2}{8M_b^3} +
  c_2\frac{ig}{8M_b^2}\left(\boldmath\Delta\cdot {\bf E} -
  {\bf E}\cdot\boldmath\Delta\right) \nonumber \\
  && - c_3 \frac{g}{8M_b^2}\sigma\cdot\left( \tilde\Delta\times\tilde {\bf E} -
  \tilde{\bf E}\times\tilde\Delta\right) \nonumber \\
  && - c_4\frac{g}{2M_b}\sigma\cdot\tilde{\bf B} \nonumber \\
  && + c_5 \frac{a^2\Delta^{(4)}}{24M_b} - c_6
  \frac{a\left(\Delta^{(2)}\right)^2}{16nM_b^2} \; ,
\end{eqnarray}
and spin-sensitive corrections of order $M_bv^6$, 
\begin{eqnarray}
  \delta H^{(6)} & = & - c_7 \frac{g}{8M_b^3}\lbrace
  \Delta^{(2)},\sigma\cdot {\bf B}\rbrace \nonumber \\
  && - c_8 \frac{3g}{64M_b^4} \lbrace
  \Delta^{(2)},\sigma\cdot\left(\Delta\times{\bf E} - {\bf
  E}\times\Delta\right)\rbrace \nonumber \\
  && - c_9 \frac{ig^2}{8M_b^3}\sigma\cdot{\bf E}\times {\bf E} \; .
\end{eqnarray}
The ${\bf E}$ and ${\bf B}$ fields are represented by the standard clover term or,
where appropriate, by an improved version to remove ${\cal O}\left(a^2
  M_b v^4\right)$ discretization errors.
Following \cite{davies} the quark Greens functions are calculated from the
evolution equation 
\begin{eqnarray}
  G(t+1) & = & \left( 1 - \frac{aH_0}{2n}\right)^n U_4^{\dag}\left( 1 -
  \frac{aH_0}{2n}\right)^n\nonumber \\
  && \times \left( 1 - a\delta H \right) G(t), \\
  G(0) & = & \delta_{{\bf x},0} \; , \nonumber
\end{eqnarray}
with $n=2$ \cite{lepage}. We rely on tapole
improvement choosing the mean link in Landau gauge as our improvement
scheme. The coefficients $c_i$ are then set to their tree level value
of 1. 
%
%
\subsection{Lattice Parameters}
The lattice parameters we have used are displayed in table
\ref{simulation}.
Details concerning 
the generation of our dynamical configurations and issues surrounding
autocorrelations are discussed in ref.~\cite{sesamauto}. 
We exploit configurations more than once
by starting the propagator evolution both at different spatial source
points and on different timeslices. A binning procedure confirms that
our 4 measurements per configuration are indeed independent.
Throughout, we fix the bare heavy quark mass to a value $aM_b =
1.7$.
\begin{table}
\caption{Simulation details\label{simulation}}
\begin{center}
\begin{tabular}{cccc}
\hline
\multicolumn{4}{c} {$\beta_{\rm dyn} = 5.6$, $n_f = 2$, $16^3 \times
  32$}\\
\hline
$\kappa_{\rm sea}$ & 0.156 & 0.1570 & 0.1575 \\ 
\hline
\# configurations & 200 & 200 & 200 \\   
\hline
measurements & 800 & 800 & 800 \\
\hline
\hline
\multicolumn{4}{c} {$\beta_{\rm quenched} = 6.0$, $n_f = 0$, $16^3 \times 32$}\\
\hline
\multicolumn{4}{c}{\# configurations: 200}\\
\hline
\multicolumn{4}{c}{measurements: 800}\\
\hline
\end{tabular}
\end{center}
\vspace{-1.0cm}
\end{table}
\begin{figure}[htb]
\begin{center}
\noindent\parbox{7.5cm}{
\parbox{7cm}{\epsfxsize=7cm\epsfysize=4cm\epsfbox{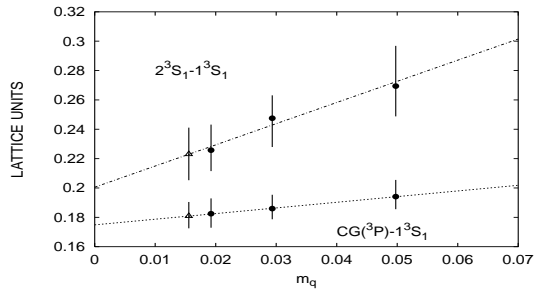}}}
\end{center}
\vspace{-1.4cm}
\caption{Extrapolation of spin independent splittings to
  $m_s/3$.The triangular symbol denotes the
  extrapolated value.\label{chiral1}}
\vspace{-1cm}
\end{figure}
%
%
\section{SPECTROSCOPY}
For details concerning the choice of operators, smearing techniques
and fitting procedures we refer to \cite{sesamnrqcd}.
We extrapolate energy level splittings linearly in the quark mass:
\begin{equation}
\Delta m = \Delta m_0  + c \sum_{u,d,s} m_q  \, ,
\end{equation}
where $m_q$ denotes the bare light quark masses(see figure
\ref{chiral1} for an example). The relevant scale we pick is
given by $\frac{m_u + m_d + m_s}{3} \approx \frac{m_s}{3}$.
The value of the strange quark mass is taken from our
recent light quark mass calculation \cite{sesamlight} and is such that
$\frac{m_s}{3}\simeq 0.0156$ lies close to our lightest sea quark mass.
\begin{table}
\begin{center}
\begin{tabular}{ccc}
\hline
& Average $a^{-1}$[{\sf GeV}] & $R_{\sf SP}$ \\
\hline\hline
$n_f = 0$, $\beta = 6.0$ & 2.47(10) & 1.43(9) \\
\hline
$\kappa = 0.1560 $ & 2.18(12) & 1.39(13) \\
\hline 
$\kappa = 0.1570 $ & 2.32(10) & 1.33(12) \\
\hline 
$\kappa = 0.1575 $ & 2.45(12) & 1.24(13) \\
\hline
$ m_s/3 $ & 2.48(14) & 1.23(11) \\
\hline
\end{tabular}
\vspace{0.1cm}
\caption{We use the average lattice spacing from
  the $2^3S_1-1^3S_1$ and $1^3\bar P-1^3S_1$ to convert our
  results to physical units. $R_{\sf SP}$ is to be compared to the
  experimental value of 1.28.\label{tab_spacings}}
\end{center}
\vspace{-1cm}
\end{table}
%
%
%
%
%
\begin{figure}[tb]
\setlength{\unitlength}{.016in}
\begin{minipage}{6.5cm}
\begin{center}
\begin{picture}(130,140)(10,930)
\put(15,935){\line(0,1){125}}
\multiput(13,950)(0,50){3}{\line(1,0){4}}
\multiput(14,950)(0,10){10}{\line(1,0){2}}
\put(12,950){\makebox(0,0)[r]{9.5}}
\put(12,1000){\makebox(0,0)[r]{10.0}}
\put(12,1050){\makebox(0,0)[r]{10.5}}
\put(12,1070){\makebox(0,0)[r]{GeV}}
\put(15,935){\line(1,0){150}}


\put(40,930){\makebox(0,0)[t]{${^1S}_0$}}



%

\put(25, 942.9){\circle{2}}
\put(25, 942.9){\line(0,1){0.1}}
\put(25, 942.9){\line(0,-1){0.1}}

\put(25,1002.8){\circle{2}}
\put(25,1002.8){\line(0,1){2.1}}
\put(25,1002.8){\line(0,-1){1.6}}


\put(35,1001.8){\circle*{2}}
\put(35,1001.8){\line(0,1){2.6}}
\put(35,1001.8){\line(0,-1){2.4}}

\put(35, 942.7){\circle*{2}}
\put(35, 942.7){\line(0,1){0.2}}
\put(35, 942.7){\line(0,-1){0.1}}


\put(80,930){\makebox(0,0)[t]{${^3S}_1$}}
\put(90,940){\makebox(0,0){1S}}
\multiput(65,946)(3,0){9}{\line(1,0){2}}
\put(90,990){\makebox(0,0){2S}}
\multiput(65,1002)(3,0){9}{\line(1,0){2}}
\put(90,1026){\makebox(0,0){3S}}
\multiput(65,1036)(3,0){9}{\line(1,0){2}}

\put(70,946){\circle{2}}
\put(70,1005.8){\circle{2}}
\put(70,1005.8){\line(0,1){1.6}}
\put(70,1005.8){\line(0,-1){2.0}}

\put(70,1061.4){\circle{2}}
\put(70,1061.4){\line(0,1){7.5}}
\put(70,1061.4){\line(0,-1){7.5}}

\put(75,946){\circle*{2}}
\put(75,1001.3){\circle*{2}}
\put(75,1001.3){\line(0,1){2.5}}
\put(75,1001.3){\line(0,-1){2.4}}

\put(75,1042.3){\circle*{2}}
\put(75,1042.3){\line(0,1){11.6}}
\put(75,1042.3){\line(0,-1){16.2}}


\put(120,930){\makebox(0,0)[t]{$^1P_1$}}

\put(140,990){\makebox(0,0){1P}}
\multiput(100,990)(3,0){9}{\line(1,0){2}}
\put(140,1026){\makebox(0,0){2P}}
\multiput(100,1026)(3,0){9}{\line(1,0){2}}

\put(105, 987.6){\circle{2}}
\put(105, 987.6){\line(0,1){1.3}}
\put(105, 987.6){\line(0,-1){0.9}}

\put(105,1039.6){\circle{2}}
\put(105,1039.6){\line(0,1){7.3}}
\put(105,1039.6){\line(0,-1){5.6}}

\put(115, 991.0){\circle*{2}}
\put(115, 991.0){\line(0,1){2.2}}
\put(115, 991.0){\line(0,-1){2.0}}

\put(115,1031.2){\circle*{2}}
\put(115,1031.2){\line(0,1){9.1}}
\put(115,1031.2){\line(0,-1){7.1}}

\end{picture}
\end{center}
\end{minipage}
\vspace{-0.5cm}
\caption{
The $b\bar b$ spectrum:radial and angular momentum
splittings with the $\Upsilon$-level set to its 
physical value. Open circles : $n_f = 0, \beta = 6.0$; filled
circles : $n_f = 2, m_q = m_s/3$; lines: experiment.\label{fig:spectrum}}
\vspace{-0.5cm}
\end{figure}
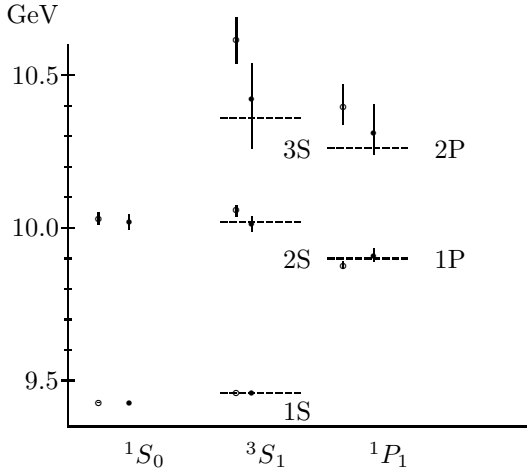

\begin{figure}
%
%
%
%
\begin{center}
\setlength{\unitlength}{.017in}
\begin{minipage}{5.7cm}
\begin{picture}(100,80)(15,-50)
\put(15,-50){\line(0,1){80}}
\multiput(13,-40)(0,20){4}{\line(1,0){4}}
\multiput(14,-40)(0,10){7}{\line(1,0){2}}
\put(12,-40){\makebox(0,0)[r]{$-40$}}
\put(12,-20){\makebox(0,0)[r]{$-20$}}
\put(12,0){\makebox(0,0)[r]{$0$}}
\put(12,20){\makebox(0,0)[r]{$20$}}
\put(12,40){\makebox(0,0)[r]{MeV}}
\put(15,-50){\line(1,0){130}}

\multiput(28,0)(3,0){9}{\line(1,0){2}}
\put(60,2){\makebox(0,0)[t]{$\Upsilon$}}
\put(60,-34){\makebox(0,0)[t]{$\eta_b$}}


\put(35,0){\circle{2}}
\put(35, -31.3){\circle{2}}
\put(35, -31.3){\line(0,1){1.2}}
\put(35, -31.3){\line(0,-1){1.0}}

\put(45,0){{\makebox(0,0){\circle*{2}}}}
\put(45, -33.4){\circle*{2}}
\put(45, -33.4){\line(0,1){1.9}}
\put(45, -33.4){\line(0,-1){1.5}}

\multiput(68,0)(3,0){10}{\line(1,0){2}}
\put(100,0){\makebox(0,0)[l]{$h_b$}}

\put(73,  -2.8){\circle{2}}
\put(73,  -2.8){\line(0,1){1.7}}
\put(73,  -2.8){\line(0,-1){2.1}}

\put(83,   0.9){\circle*{2}}
\put(83,   0.9){\line(0,1){2.9}}
\put(83,   0.9){\line(0,-1){2.8}}

\multiput(110,-40)(3,0){10}{\line(1,0){2}}
\put(140,-40){\makebox(0,0)[l]{$\chi_{b0}$}}
\multiput(110,-8)(3,0){10}{\line(1,0){2}}
\put(140,-8){\makebox(0,0)[l]{$\chi_{b1}$}}
\multiput(110,13)(3,0){10}{\line(1,0){2}}
\put(140,13){\makebox(0,0)[l]{$\chi_{b2}$}}

\put(115, -27.1){\circle{2}}
\put(115, -27.1){\line(0,1){4.1}}
\put(115, -27.1){\line(0,-1){3.8}}

\put(115,  -6.7){\circle{2}}
\put(115,  -6.7){\line(0,1){2.3}}
\put(115,  -6.7){\line(0,-1){3.2}}

\put(115,   9.4){\circle{2}}
\put(115,   9.4){\line(0,1){1.7}}
\put(115,   9.4){\line(0,-1){2.5}}

\put(125, -28.3){\circle*{2}}
\put(125, -28.3){\line(0,1){6.9}}
\put(125, -28.3){\line(0,-1){6.3}}

\put(125,  -4.0){\circle*{2}}
\put(125,  -4.0){\line(0,1){3.7}}
\put(125,  -4.0){\line(0,-1){2.7}}

\put(125,   8.0){\circle*{2}}
\put(125,   8.0){\line(0,1){3.1}}
\put(125,   8.0){\line(0,-1){2.7}}

\end{picture}
\end{minipage}
\end{center}
\vspace{-0.8cm}
\caption{
Fine structure: here the
zero of energy is set to the $\Upsilon$-level in the left part and to
the spin averaged triplet P -level in the right part. Labels as in \ref{fig:spectrum}.
\label{fig:finestructure}}
\vspace{-0.8cm}
\end{figure}
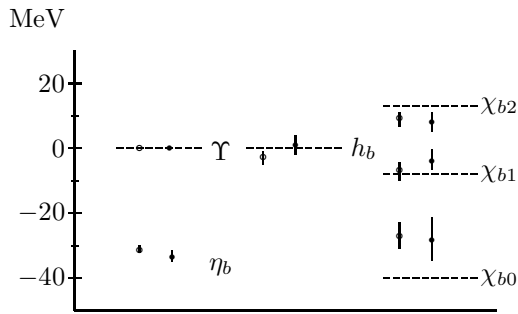

%
Figure \ref{fig:spectrum} and \ref{fig:finestructure} summarize our
results for 2 and 0 flavours. Unquenching effects are clearly
visible in the spin-independent part of the $b \bar b$ spectrum. This
is also evident from table \ref{tab_spacings}:
$R_{SP}\equiv (2^3S_1-1^3S_1)/(1^3\bar P-1^3S_1)$ 
disagrees with the experimental value $R_{SP}=1.28$ in the quenched
case but coincides with it when dynamical fermions are included.
%

%
\par
We do not observe any significant impact of unquenching on the hyperfine
splittings. In particular, the P-hyperfine splittings seem to be
underestimated for both $n_f=0$ and $n_f=2$. Clearly, this result
needs to be corroborated with higher statistics. Errors on our
hyperfine splittings encompass the statistical error as well as the
uncertainty in the fitting range; the latter is essentially of the
same size as the statistical error so that we can expect some improvement
with higher statistics.
\par
\begin{figure}[htb]
\begin{center}
\noindent\parbox{7.5cm}{
\parbox{7cm}{\epsfxsize=7cm\epsfysize=3.0cm\epsfbox{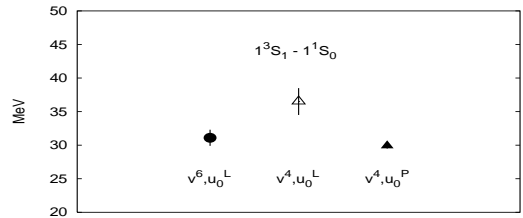}}}
\end{center}
\vspace{-1.2cm}
\caption{Hyperfine S-splitting obtained
  with an ${\cal O}(M_b v^6)$ and ${\cal O}(M_b v^4)$ correct NRQCD
  action. The higher order result is 
  shifted significantly downward. Also shown: result with an ${\cal
  O}(M_b v^4)$ action but using the plaquette tadpole scheme. 
  All results are for the quenched case with $\beta = 6.0$.
\label{figsystematics}}
\vspace{-1.2cm}
\end{figure}

We expect the inclusion of ${\cal
  O}(M_b v^6)$ terms to effect the hyperfine splittings on the 10 \%
level, since $v^2 \approx 0.1$. Performing a relatively inexpensive ${\cal
  O}(M_b v^4)$ quenched simulation at $\beta = 6.0$ (with 200
measurements) we find this naive expectation
to be well satisfied for the $^3S_1 - {^1S_0}$ splitting as is
shown in figure \ref{figsystematics}. However, using the
  plaquette prescription for $u_0$ (result from the NRQCD-Collab.,
  also included in figure \ref{figsystematics}) shifts the value on
  the 10\% level in
  the opposite direction as adding ${\cal O}(M_bv^6)$ corrections. 
%
%
\section{SUMMARY AND OUTLOOK}
Overall, NRQCD has proven to work very well in the spin-independent
sector, in particular in giving a precise determination of the lattice
spacing $a$. It appears worthwhile to push the scale determinations to higher
statistics using our configurations generated by \tkl at $\kappa =
0.1575$ on $24^3 \times 40$ lattices ($\beta = 5.6$). This would
enable a reliable extrapolation to 3 or 4 flavours. The dominant
source of error in such a lattice scale determination is most likely
due to the remaining lattice discretisation errors in the gauge
configurations. 
\par
Although higher statistics are highly desirable, it seems unlikely that
progress in the spin-dependent sector will come from this
approach alone. It is worrying to find the spin-dependent corrections of
${\cal O}(M_b v^6)$ as large as 10 \% even for the $b
\bar b$, decreasing the fine-splittings relative to the lower
order estimates. A perturbative or
non-perturbative calculation of the coefficients $c_i$ is badly
needed.

\end{document}